\documentclass[fleqn,twoside]{article}
\usepackage{espcrc2}

\usepackage{graphicx}
\usepackage[figuresright]{rotating}

\newcommand{\AmS}{{\protect\the\textfont2
  A\kern-.1667em\lower.5ex\hbox{M}\kern-.125emS}}

\title{Zero temperature phase structure of multi-flavor QCD\thanks{Presented by S.K.}}

\author{Seyong Kim\address[MCSD]{Department of Physics, 
        Sejong University, Seoul, Korea}
        \thanks{S.K. is supported by Korea Research Foundation}
	and
        Shigemi Ohta\address{Institute for Particle and Nuclear Studies, 
	KEK, Japan / RIKEN-BNL Reseaerch Center, BNL, USA}}
       
\begin{document}

\begin{abstract}
To study continuum limit of lattice QCD with many light quark flavors,
we investigate the zero temperature phase structures of multi-flavor
QCD.  Currently a series of exploratory simulations are being performed with the
number of quark flavors, $N_f $= 6,  8, and 10 on a set of lattice
volumes for various quark masses.  Here, we report on the current
status of our simulation on a $8^4$ lattice.
\vspace{1pc}
\end{abstract}

\maketitle

\section{Introduction}

Perturbative analysis of QCD coupling constant renormalization tells
us that the coefficients in the series expansion of the
$\beta$-function changes their signs depending on the number of light
quark flavors. Based on this, it was suggested that long and/or
short distance behavior of QCD may change with the number of light
quark flavors \cite{Banks_Zaks}. Although these arguments are founded
upon perturbation theory, they may remain valid even in the
non-perturbative regime. Indeed, lattice simulations with
multi-flavored QCD reveal rich phase
structures\cite{Kogut_et_al,Columbia,CP-PACS}. 

For eight light staggered quark flavors, there is a strong first order
phase transition which separates the strong coupling region from the
weak coupling region. For a given lattice spatial volume, this
transition appears to be $N_t$-independent bulk transition where $N_t$
is the number of sites along the time direction. The weak coupling
phase is divided into two region : in one region of weak coupling
space chiral condensates shows linear behavior in quark mass, and in
the other region they show non-linearity\cite{Columbia}. It has been
speculated that there may be a normal finite temperature phase
transition between these two different weak coupling phases. On the
other hand, using simulation results with Wilson quark formulation in
the strong coupling limit, the authors of Ref.\ \cite{CP-PACS} argue
that lattice QCD for $N_f \ge 7$ has an interacting limit without
quark confinement in contrast to the usual QCD, a theory with
spontaneous chiral symmetry breaking and color confinement in low
energy. Even if the color confinement and the spontaneous chiral
symmetry breaking is rigorously proven in the infinite coupling limit
(in quenched approximation to QCD), it is claimed that copious
addition of light quark flavors modifies the string-like vacuum
structure of a theory with gluon only.

In short, the existence of a strong first order bulk phase transition
which separates the strong coupling region and the weak coupling one
is well established. Continuum limit of lattice theory for
multi-flavor QCD is less clear. For staggered quark simulations,
investigation of the weak coupling phase by use of hadron spectroscopy
calculation was hampered by small spatial lattice
volumes \cite{Columbia}.  Simulation result with Wilson dynamical quark
in the strong coupling limit is difficult to make a contact with the
continuum limit. Thus, further study in the weak coupling phase
structure of multi-flavor QCD using light dynamical quarks is
needed. Here, we would like to endeavor toward this direction. Our
efforts will be concentrated on various susceptibilities such as
chiral susceptibility and finite size scaling of them. 

\section{Simulation Characteristics}

\begin{figure}[t]
\begin{center}
\includegraphics[width=60mm,bb=161 250 452 552]{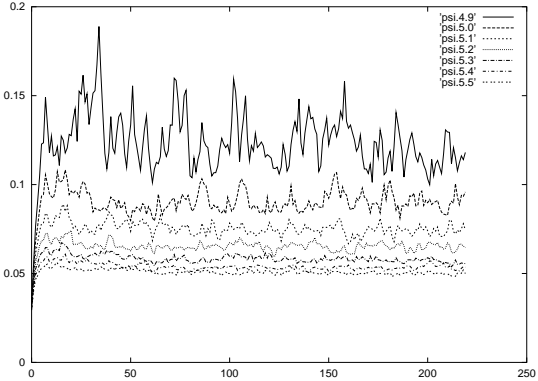}
\caption{Monte Carlo time evolution of chiral condensate for $N_f = 6$ 
with $m_q a = 0.01$.}
\label{fig:evolution}
\end{center}
\end{figure}

We use R-algorithm\cite{Gottlieb} to simulate QCD with arbitrary
number of quark flavors and employ staggered fermion method for the
dynamical quarks. Target platform for our project is 128-node Fujitsu
VPP-700 at RIKEN, the Institute of Physical and Chemical Research, in
Japan. Since we expect the need for large lattice volume from other
groups' experiences, two dimensional layout of compute nodes is
adopted as part of our parallelization strategy. Similar to our work
on quenched spectroscopy of QCD\cite{VPP700}, four dimensional $N_t
\times N_x \times N_y \times N_z$ lattice points are distributed
evenly over the y and the z directions so that for given number of
compute nodes, it is divided into $M_y \times M_z $'s y directional
and z directional nodes. Each compute nodes handles $N_t \times N_x
\times L_y \times L_z $ sub-lattice points. For example, $8^4$ lattice
volume run reported here uses 8-node partition. This 8-node partition
is layed out as $(4,2)$ mesh points and each nodes computes $8\times 8
\times 2 \times 4$ sub-lattice points.  We also use checker-boarded
site classification for storing dynamical variables. Periodic boundary
condition for the time direction is imposed on the gauge
field. Anti-periodic boundary condition for the time direction is
imposed on the fermion. For the three space directions, periodic
boundary condition is used for all fields.
\begin{figure}[t]
\begin{center}
\includegraphics[width=60mm,bb=161 250 452 552]{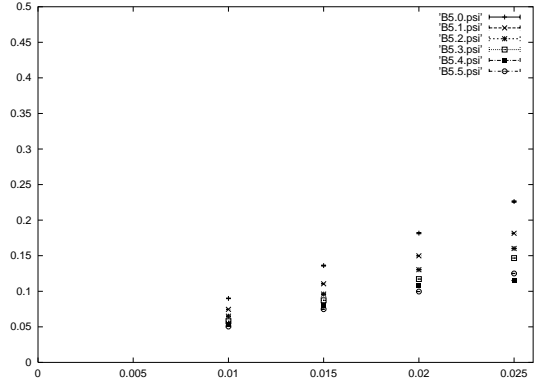}
\caption{Chiral condensate for $N_f = 6$ vs $m_q a = 0.01, 0.015,
0.02, 0.025$ at different $\beta$'s}
\label{fig:linearity}
\end{center}
\end{figure}
Before studying zero temperature phase structure of multi-flavor QCD
in detail, a set of exploratory runs on a $8^4$ lattice with time step $\Delta
\tau = 0.01$ was done. Simulations with dynamical quark mass $m_q a =
0.01, 0.015, 0.02$, and $0.025$ is performed for the number of quark
flavors, $N_f = 6$ and 8. Since we are interested in the weak coupling
phase, range of $\beta = 6/g^2$ from 5.0 to 5.5 is chosen for the
preliminary run (note that for a $16^4$ lattice volume with $N_f = 8$
staggered quarks, the critical coupling is 4.62(1) for $\Delta \tau =
0.005$\cite{Columbia}). There is $(\Delta \tau)^2$ errors associated
with R-algorithm and the difference in the step size makes
quantitative comparison more subtle.

\begin{figure}[t]
\begin{center}
\includegraphics[width=60mm,bb=161 250 452 552]{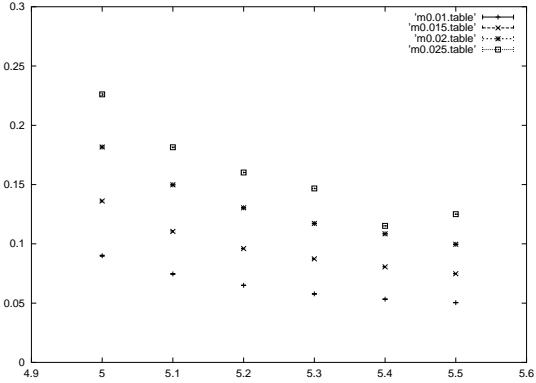}
\caption{Chiral condensate for $N_f = 6$ for each $m_q a$ at different coupling constant, $\beta = 6/g^2$}
\label{fig:mass6}
\end{center}
\end{figure}

\begin{figure}[t]
\begin{center}
\includegraphics[width=60mm,bb=161 250 452 552]{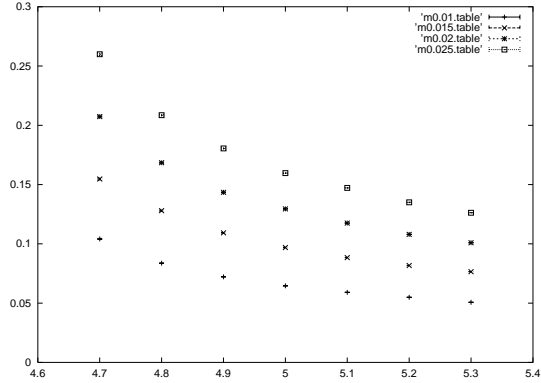}
\caption{Chiral condensate for $N_f = 8$ for each $m_q a$ at different coupling constant, $\beta = 6/g^2$}
\label{fig:mass8}
\end{center}
\end{figure}

Figure \ref{fig:evolution} shows typical Monte Carlo evolutions of the chiral condensate
for four different quark masses with $N_f = 6$ at each $\beta$.
Ordered start runs with total simulation time, $\tau \sim 220$, are shown in
the figure. The average values of the chiral condensate from the Monte Carlo time
$\tau = 20$ to $\tau = 220$ are given in Figure \ref{fig:mass6}.  The error in the
figure does not include auto-correlation among the data and are
probably underestimated.  There is a linear relationship between the
chiral condensate and the quark mass and it is in agreement with the
behavior found in \cite{Columbia}.
Similarly, the average chiral condensate for $N_f = 8$ is shown in Figure \ref{fig:mass8}. The simulation length, $\tau$, is $\sim 220$ and the average is over the last $\tau = 200$
time unit. Ranges of $\beta = 4.7 \sim 5.5$ are simulated with the
same set of quark masses as $N_f = 6$.

\section{Discussion}

In order to study the zero temperature behavior of multi-flavor QCD
further, we started full dynamical simulation. R-algorithm which
allows us to simulate arbitrary number of quark flavors is chosen. Our
simulation is at early stage of investigation and is currently
compared with earlier results by other groups on a small $8^4$ lattice
volume. The weak coupling phase of $N_f = 6$ and 8 is
investigated. There exist a linear relation between the chiral
condensate and the light dynamical quark mass, which is in agreement
with earlier results. For the planned more detailed investigation of
the weak coupling phase, several aspects of the current lattice
calculation need to be refined.  First, we need to test whether
$\Delta \tau^2$ error associated with the current algorithm is small
in the simulation result. Next, we need to add more observables :
various susceptibilities such as chiral susceptibility calculation and
specific heat calculation are currently being implemented.

\end{document}